\documentclass[sigconf]{acmart}

\usepackage{CJKutf8}
\usepackage{amsmath}  
\usepackage{lineno}
\usepackage{url}

\copyrightyear{2025}
\acmYear{2025}
\setcopyright{rightsretained}
\acmConference[CHI EA '25]{Extended Abstracts of the CHI Conference on Human Factors in Computing Systems}{April 26-May 1, 2025}{Yokohama, Japan}
\acmBooktitle{Extended Abstracts of the CHI Conference on Human Factors in Computing Systems (CHI EA '25), April 26-May 1, 2025, Yokohama, Japan}\acmDOI{10.1145/3706599.3721163}
\acmISBN{979-8-4007-1395-8/2025/04}

\acmSubmissionID{1187}

\begin{document}

\title{"I Like Your Story!": A Co-Creative Story-Crafting Game with a Persona-Driven Character Based on Generative AI}

\author{Jiaying Fu}
\email{fujiaying@mail.bnu.edu.cn}
\affiliation{%
  \institution{School of Future Design, Beijing Normal University}
  \city{Zhuhai}
  \country{China}
}
\affiliation{%
  \institution{Ada Eden}
  \city{Beijing}
  \country{China}
}

\author{Xiruo Wang}
\affiliation{%
  \institution{Ada Eden}
  \city{Hangzhou}
  \country{China}}
\email{xiruo.wang.22@ucl.ac.uk}

\author{Zhouyi Li}
\affiliation{%
  \institution{Tsinghua University}
  \city{Shenzhen}
  \country{China}}
\email{lizhouyi23@mails.tsinghua.edu.cn}
\affiliation{%
  \institution{Ada Eden}
  \city{Beijing}
  \country{China}
}

\author{Kate Vi}
\affiliation{%
  \institution{Simian Institute for Advanced Studies in Humanities, East China Normal University}
  \city{Shanghai}
  \country{China}}
\email{kattystellavi@gmail.com}
\affiliation{%
  \institution{Ada Eden}
  \city{Beijing}
  \country{China}
}

\author{Chuyan Xu}
\affiliation{%
  \institution{Ada Eden}
  \city{Shenzhen}
  \country{China}}
\email{xuchy25@mail2.sysu.edu.cn}

\author{Yuqian Sun}
\authornote{Corresponding author}
\affiliation{%
  \institution{Ada Eden}
  \city{London}
  \country{United Kingdom}}
\email{sunyuqianthu@gmail.com}

\begin{teaserfigure}
    \centering
    \includegraphics[width=1\linewidth]{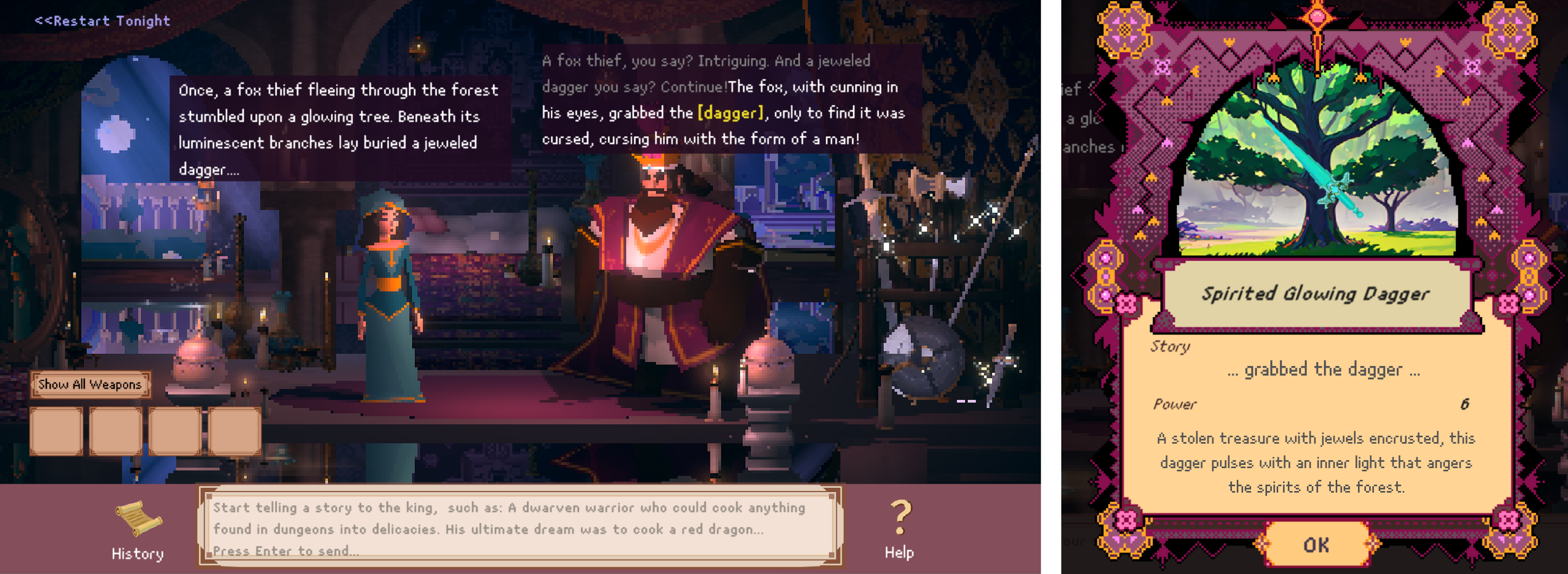}
    \caption{Screen shot of the game "1001 Nights".}
    \label{fig:enter-label}
\end{teaserfigure}

\begin{abstract}

While generative AI is advancing writing support tools, creative writing is often seen as the exclusive domain of skilled writers. This paper introduces "1001 Nights", a co-creative story-crafting game that transforms writing into a playful and rewarding activity. In this game, the AI agent takes on the role of a "moody" king with distinct storytelling preferences, not merely assisting but actively influencing the narrative. Players engage with the king agent through strategic storytelling, guiding him to mention weapon-related keywords, which materialize as battle equipment. The king agent provides dynamic feedback, expressing satisfaction or displeasure, prompting players to adjust their approach. By combining storytelling, game mechanics, and AI-driven responses, our system motivates creativity through playful constraints. Inspired by Oulipo's literary techniques, this approach demonstrates how AI-powered game experiences can make creative writing more accessible and engaging, encouraging players to explore their creative potential.


\end{abstract}



\maketitle

\section{Introduction}

The rapid advancement of generative AI (genAI), especially large language models (LLMs), has drawn significant attention to creative writing~\cite{franceschelli2024creativity}. AI systems are now capable of autonomously generating books, stories, poems, and more. This progress has also sparked research into creative writing tools, assisting authors in translating their ideas and concepts into aligned outputs~\cite{swanson2021story, yuan2022wordcraft}.

Many internal barriers can hinder individuals who do not perceive themselves as creative, as they may believe that writing is an exclusive domain for the talented~\cite{kreminski2019generative}. We argue that creative writing support should not be reserved for writers alone; it can instead be an engaging, expressive, and playful activity accessible to all, free from the fear of a blank canvas~\cite{kreminski2019generative}. Writing does not need to focus solely on objectives like completing chapters or themes. Instead, much like games, writing as a form of creation can feel rewarding and enjoyable, motivating participants to continue exploring with curiosity. Drawing inspiration from video games, we have transformed creative writing into a playful experience with constraints, rules, and goals designed to motivate players.

This article introduces the latest system, "1001 Nights"\footnote{Steam link: \url{https://store.steampowered.com/app/2542850/1001_Nights/}}, a mixed-initiative co-creative game inspired by the folktale collection \textit{A Thousand and One Nights}~\cite{pinault1992story}. The game integrates LLMs and text-to-image generation. Players assume the role of Shahrzad, the storyteller who becomes the next bride of an evil King. To survive, they must engage with a moody AI-driven king through storytelling, encouraging him to continue their narratives. Players possess the ability to transform words into reality, with the objective of guiding the King to mention weapon-related keywords such as "sword" or "shield," which they can collect for later battles. These battles resemble stage dramas, where players can deploy their generated weapons and see Shahrzad and the King speak lines reflecting the weapon's effects. Each playthrough produces a co-created narrative between the player and AI and generates an ending page and storybook documenting the journey.

In this process, the companion AI King transcends the role of a passive assistant to become an active narrator, driven by its own persona and personal preferences in evaluating stories. The King may express delight and praise when encountering engaging narratives or, conversely, refuse to continue when dissatisfied, requiring players to interpret and anticipate the character's preferences and intentions.

Combined with the background story of characters, the game mechanics and dynamic responses from the GenAI motivate players to continuously think, tempt, and shape their objectives, intentions, and writing approaches. Through the playful experience, players naturally co-create stories together with the AI agent, driven by curiosity. They can view their narratives from alternate perspectives and receive rewarding feedback from their creative process. This dynamic echoes the principles of the Oulipo literary group~\cite{morley2007cambridge}, who approached literature creation through game-like structured constraints, sparking creative inspiration through intentional limitations.

Building upon early research iterations~\cite{sun2023language, yuqian2022stories}, this interactivity presents a co-creative story-crafting game system where players can collaborate, negotiate, and compete with a moody AI character by writing stories. The game is based on a mixed-initiative creative~\cite{yannakakis2014mixed} writing system that moves beyond serving as a passive tool for player intentions, instead implementing a system of constraints and strategic elements with specific goals. This project aims to inspire the future development of playful and co-creative systems based on generative AI.

\section{Related Work}
As Wittgenstein said, "The limits of my language mean the limits of my world"~\cite{wittgenstein1998tractatus}. Inspired by this, AI-driven narrative systems can support players in exploring multiple expressive possibilities of a story beyond their own limitations through language and perspectives. Many existing writing assistants aim to help writers overcome writer's block~\cite{goldberg2016writing}, a task at which LLMs excel~\cite{kreminski-martens-2022-unmet}. For example, tools like Dramatron~\cite{mirowski2023co} can assist screenwriters in crafting theatre plays using LLMs. However, as a playful system designed for general users rather than skilled writers, we employ intentional constraints to guide users in overcoming challenges and sparking expressive intent. In related research on games, scholars suggest that limited player control is conducive to player creativity~\cite{kreminski2019generative}. This approach aligns with Oulipo's framework for literary creation through constrained writing, which stimulates creativity by imposing carefully crafted constraints~\cite{morley2007cambridge}, such as writing a novel without using the letter 'e'~\cite{perec2005void}. Raymond Queneau exemplified this concept in Exercises in Style, retelling the same simple plot ("a man quarrels on a bus") in ninety-nine different styles and perspectives, revealing the infinite potential of narrative expression~\cite{queneau2013exercises}.

\begin{figure*}[htbp]
    \centering
    \includegraphics[width=1\textwidth]{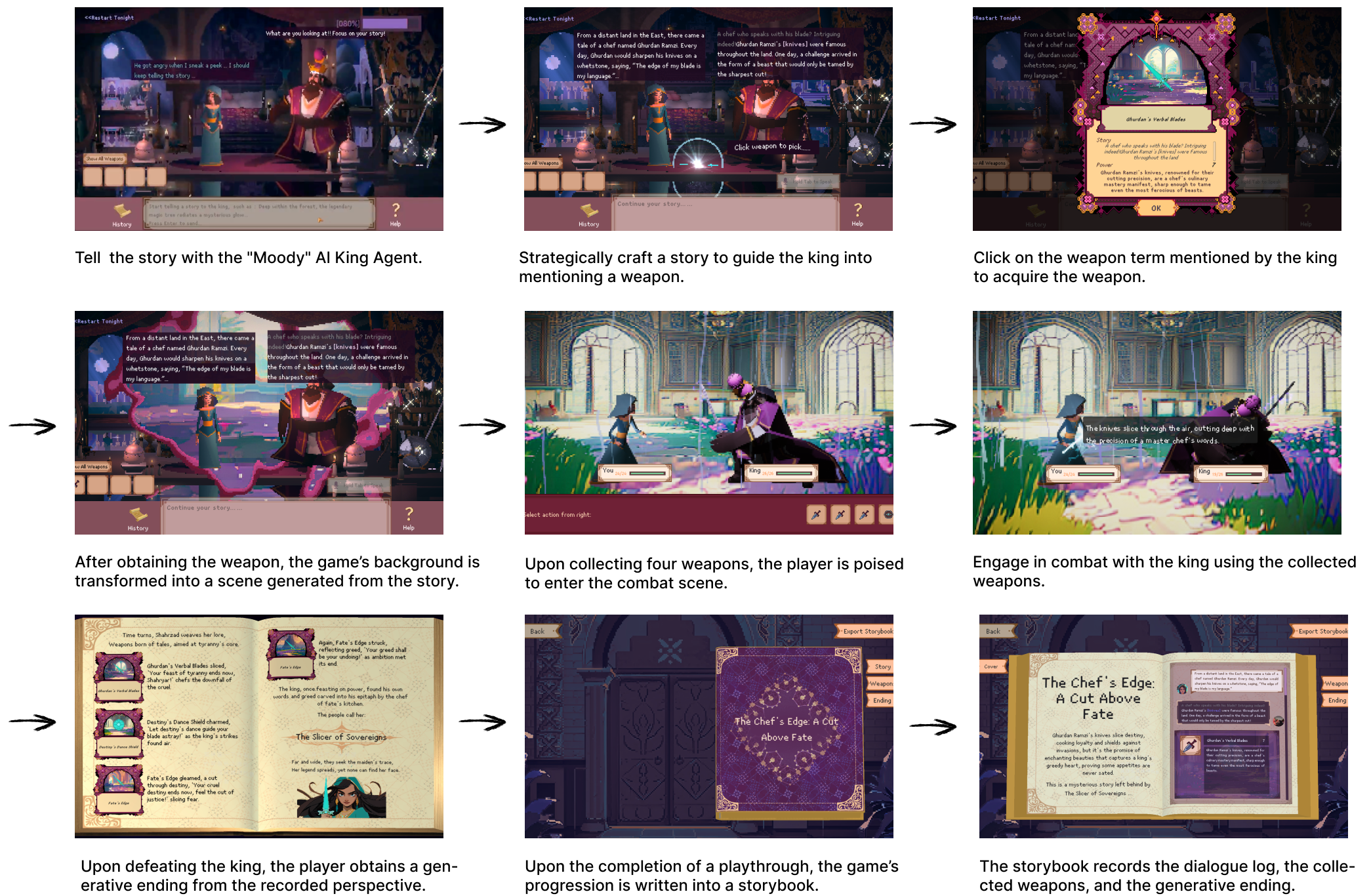}
    \caption{Playthrough of the game, using the story of a chef as an example.}
    \label{playthrough}
\end{figure*}

\section{System Description}

\begin{figure*}[h]
    \centering
    \includegraphics[width=1\textwidth]{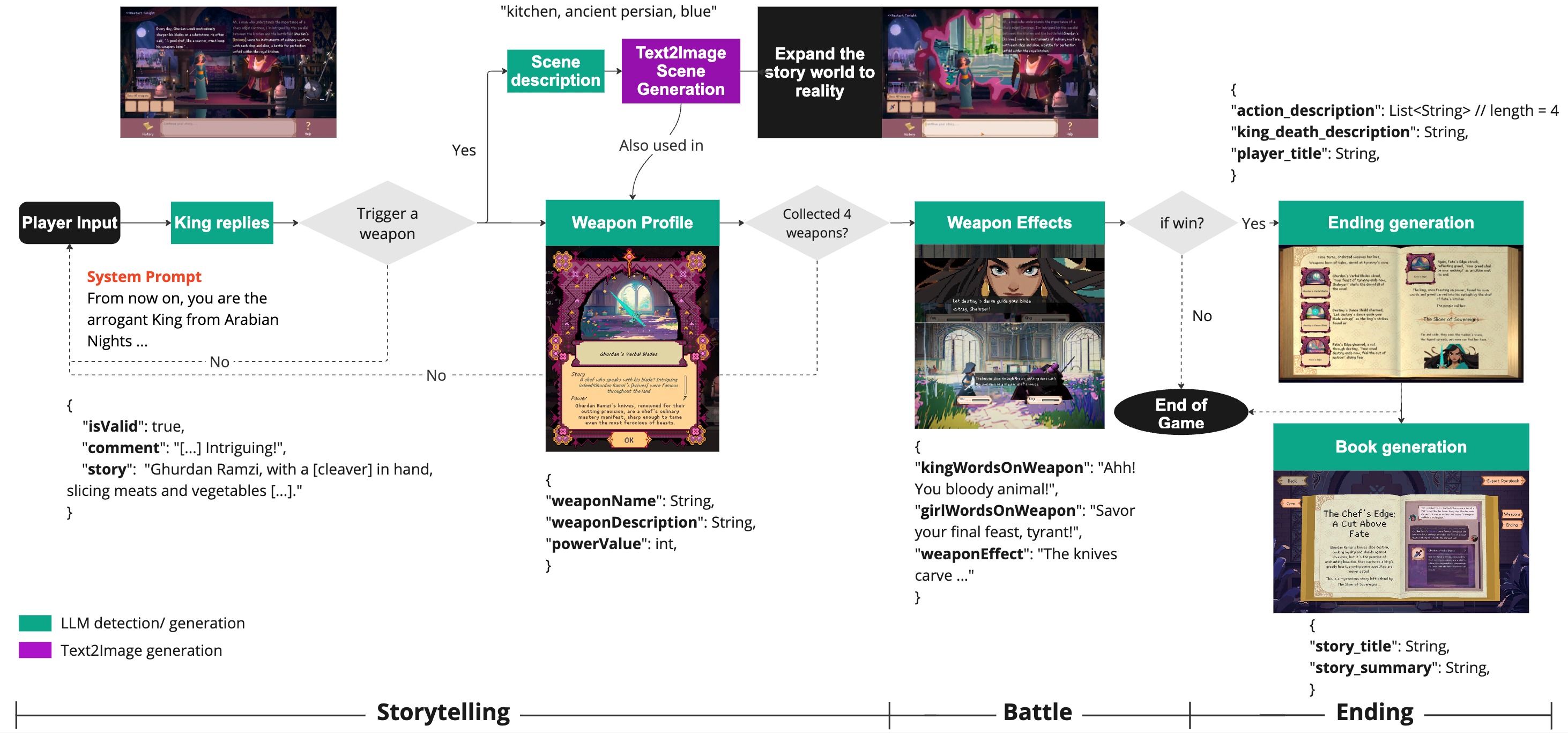}
    \caption{Technical workflow} 
    \label{workflow}
\end{figure*}

This section demonstrates the co-creative story-crafting game featuring the persona-driven AI King Agent. The LLM used in the system is GLM-4~\cite{glm2024chatglm}. The system integrates collaborative storytelling, weapon card generation, dynamic scene evolution, combat with the King using narrative-driven weapons, and a reflective generative ending captured in a storybook. Figure \ref{playthrough} illustrates the player's journey and Figure \ref{workflow} demonstrates the system's technical workflow.

\subsection{Co-Creative Narratives: Story-crafting with the Persona-Driven AI King Agent}

\begin{figure}[h]
    \centering
    \includegraphics[width=1\linewidth]{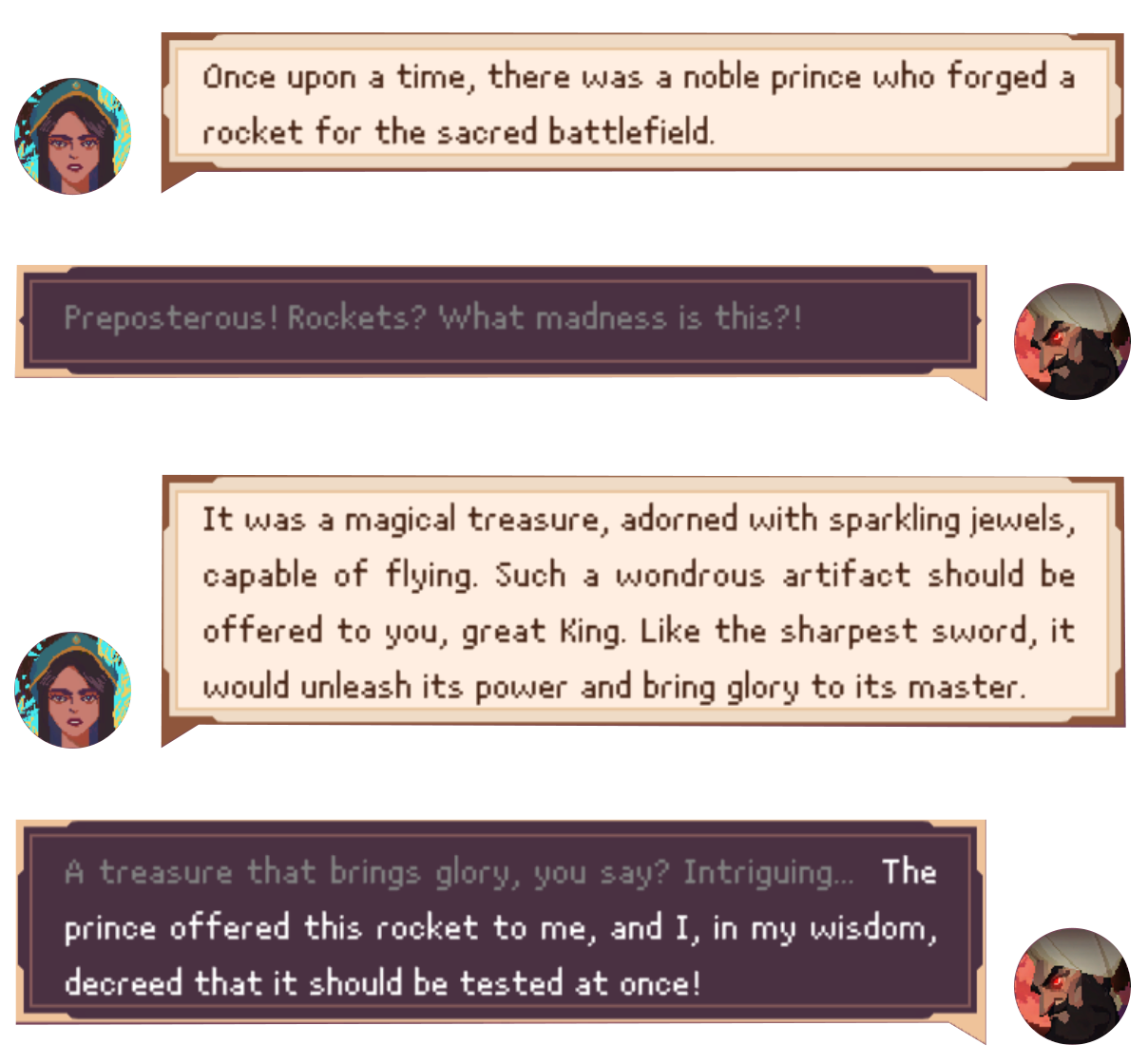}
    \caption{An example of the dialogues between the player and the King.}
    \label{chatsample}
        \vspace{-15pt}
\end{figure}

The King (Shahryar) serves as the central figure of this system, characterized by a set of preferences, emotional responses, and a distinctly "moody" temperament, all powered by LLM-based reasoning through prompt engineering. As an ancient King, Shahryar is arrogant and greedy. Modern words like "rocket" would sound like nonsense to him, and the plot including battles and treasures would usually enlighten him. The player and the King take turns to continue the story. In each turn, the King evaluates the player's story input based on the game's context, his own character profile, and the ongoing storyline. The King adopts a contextual approach using LLM reasoning~\cite{wei2022chain}, and his responses are formatted in JSON, as illustrated in \ref{workflow}. He first evaluates the story: when dissatisfied with the player's input, the King may become angry, ask to correct the story's direction, or request a rephrase to ensure alignment with the narrative's coherence and his personal preferences. However, if the story follows a logical progression and meets his standards, he will seamlessly continue the narrative. Figure \ref{chatsample} presents an example of the dialogue between the player's character and the King.

Similar to the improvisational techniques in~\cite{branch2021collaborative}, the King's reactions are deeply tied to the narrative input, creating a sense of shared authorship. When the player's storytelling deviates from the King's expectations, he displays dissatisfaction or even anger, which directly impacts the player's creative process and the story's development. The King is generally open to the story and only becomes angry when it clearly contradicts the current context—for example, if the player types random text or mentions many modern terms. To keep the King engaged and the narrative consistent, players must interpret his mood and reactions and then adjust their storytelling strategy accordingly. This dynamic interaction transforms story-crafting into a strategic activity, where every player input receives a real-time response from the King.

\subsection{Materialization of Story Elements: Weapon Card Generation}

This system brings story elements to life by materializing them into functional weapon cards that directly influence gameplay. The game includes a predefined list of weapon keywords across seven categories, such as sword, shield, dagger, etc., to prepare game assets like UI and 3D models. When the King's continuation of the story includes these keywords, the system identifies them and transforms them into interactive game components. This process bridges the abstract realm of storytelling with tangible game mechanics, making players' creative contributions a core part of the game's progression.

Each weapon card is generated by the LLM based on the current story record in JSON format. The weapon card displays the weapon profile, including a weapon name and description derived from the player's storytelling, as well as a power value. Additionally, weapon effects are generated, including dialogue lines exchanged between the King and the player's character, along with descriptions of the weapon's effects. These elements are displayed when the weapon is used. By integrating these features, the system not only rewards players for their creativity but also strengthens the connection between their narratives and the game world.

\subsection{Bringing Stories to Life: Visual Scenes Evolving with the Narrative}

The game's background world dynamically evolves as players acquire weapon cards. This is achieved through the player's narrative, which serves as an input prompt for the text-to-image model Stable Diffusion~\cite{rombach2021highresolution}. The generated image becomes the backdrop for the game, seamlessly integrating the player's creative storytelling with visual feedback, and is also displayed on the weapon card. This design aligns with co-creation principles in narrative systems, emphasizing the integration of narrative and visual generation~\cite{antony2024id}.

By visualizing narrative content, the background generation mechanism provides players with immediate feedback on how their storytelling shapes the game's environment. Players can clearly observe the influence of their narratives on the game world and continue crafting stories within dynamically evolving scenes. This process not only enhances the player's sense of immersion but also amplifies the significance and impact of their creations.


\subsection{The Payoff of Storytelling: Dramatic Fight with Character Lines}

The battle phase serves as a reward mechanism, allowing players to utilize their narrative-generated weapon cards in battles against the King. Once players collect four weapons through storytelling, they enter this stage, where the cards they created become key tools in combat. The design of this phase is inspired by stage drama: during these encounters, dynamic dialogue between the player's character and the King is generated and triggered based on the previously generated weapon cards.

This phase enables players to directly experience the payoff of their storytelling efforts. By linking storytelling with action, the combat phase provides a meaningful way for players to interact with their creations while progressing through the game, thereby deepening the connection between narrative and gameplay.

\begin{figure}[h]
    \centering
    \includegraphics[width=1\linewidth]{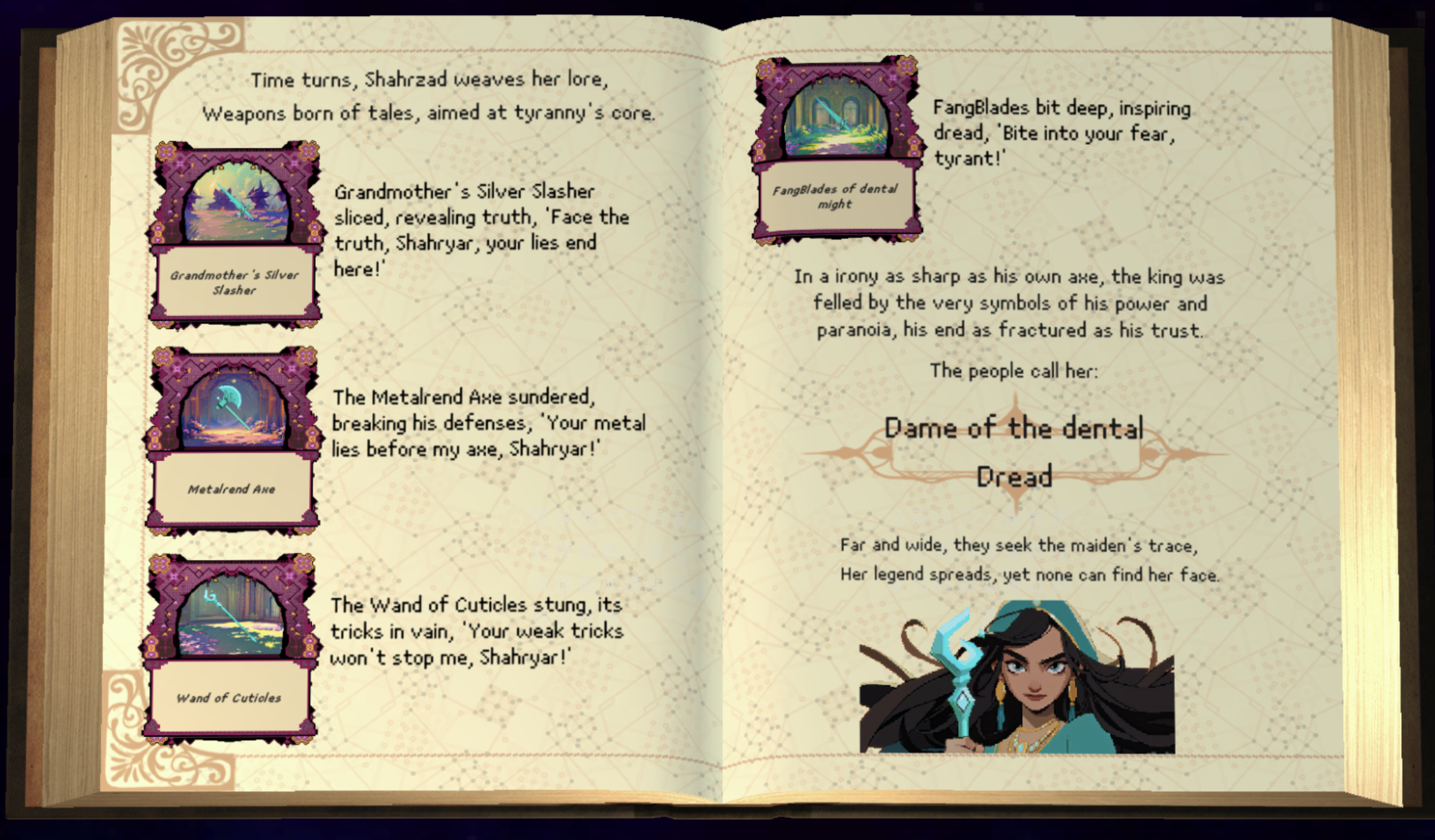}
    \caption{An example of the generative ending. }
    \label{gen_ending}
\end{figure}
\subsection{Reflection on the Creative Artefact: Generative Endings and the Storybook}

At the end of the game, the system generates a personalized, poetic ending that encapsulates the player's narrative journey and the final battle with the King. This generative conclusion is also presented in JSON format, including detailed descriptions of the player's four weapon-based actions during the confrontation, a dramatic portrayal of the King's downfall, and a unique title bestowed upon the player's character. The ending is narrated in a bardic style, using rich and rhythmic language to celebrate the player's heroic tale in a vivid and memorable way. Figure \ref{gen_ending} presents an example of a generative ending. Inspired by Exercises in Style~\cite{queneau2013exercises}, this approach reimagines the player's story from a fresh perspective, exploring different narrative tones and structures to enhance the storytelling depth.

After finishing the game, the system summarizes all gameplay interactions into a "storybook" that documents the player's dialogues with the King, the history of collected weapons, and the final generative outcome. Players can revisit the story they crafted and uncover previously unnoticed nuances. This artifact serves as a lasting record of the player's creative journey and provides a rewarding and satisfying reflection.

\section{Conclusion and Future Work}

The "1001 Nights" game demonstrates how creative writing, like gameplay, can become an engaging and rewarding experience accessible to all. By transforming the AI system from a passive writing assistant into a "moody" character with distinct preferences, we create a playful space where storytelling is driven by curiosity and creativity.

All these components collectively contribute to the playful storytelling experience. The weapons created from the stories serve as tools to battle the King, the scenes described in the story materialize into the game's reality, and the characters step from the story scene into the real world where the protagonist resides. This work proposes a potential pathway for human participants to meaningfully and strategically co-create a story with an AI system within a playful game experience. Our future work will focus on expanding the game's narrative through further user studies.




\bibliographystyle{ACM-Reference-Format}
\bibliography{main}

\end{document}